# Sub-microscopic modulation of mechanical properties in transparent insect wings

Ashima Arora, Pramod Kumar, Jithin Bhagavathi, Kamal P. Singh and Goutam Sheet[*]

Department of Physical Sciences, Indian Institute of Science Education and Research, Mohali

Punjab, India, PIN: 140306

**Abstract: We report on the measurement of mechanical properties of the transparent wings of an insect (popularly known as the "rain fly") using an atomic force microscope (AFM) down to nanometer length scales. We observe that the frictional and adhesion properties on the surface of the wings are modulated in a semi-periodic fashion. From simultaneous measurement of AFM topography it is observed that the modulation of mechanical properties is correlated with the modulation of topography. Furthermore, the regions of higher friction are decorated with finer nanostructures with definite shape. From optical diffraction experiments we show that the observed modulation and its semi-periodic nature are distributed over the entire surface of the wing.**

The ability of small insects to fly under harsh environmental conditions with apparently fragile wings has always been a matter of curiosity. With the development of high-resolution imaging capabilities, it has now been possible to investigate the detail structure of the wing-topography that is believed to play the key role in generating superior mechanical properties of the wings. [1-6]. Earlier it was revealed that certain micro-structures on the wing surface play a primary role in promoting intriguing physical properties of insect wings e.g., exotic colors, exceptionally high mechanical strength etc. [7-9]. Attempts have been made to mimic the naturally engineered structures in biological systems for enhancing the functionality of man-made devices and synthesizing new materials [3, 10-15]. In this paper we report the mechanical properties (friction and adhesion) measurement down to nanometer length scales of the transparent wings of an insect popularly known as "the rain fly" using an atomic force microscope (AFM). The acquired information might explain how the insects can fly under difficult environments (during rain and strong wind) and give new design ideas for achieving robustness through nano-structuring in artificial devices.

The frictional force measurement is done by bringing the tip, mounted at the end of a cantilever, in contact with the sample surface -- as the tip scans in contact mode it rubs against the surface and the friction leads to a lateral deflection of the cantilever which is detected optically. The lateral deflection of the cantilever is schematically described in Figure 1(a). Figure 1(b) shows a hypothetical surface with different topographic features having different coefficients of friction and the corresponding lateral deflection.  The spikes in the lateral scan lines are seen due to a sudden rise or a sudden fall in the topography. However, if there is a sudden rise(fall) in topography and the corresponding friction also rises (falls) leading to an increase(decrease) in lateral deflection in the next step, such spikes will not appear. During these, no lateral force acts on the tip. Therefore, the friction data measured during a single scan direction are always influenced by the topographic features. In our measurements we eliminate this crosstalk by recording the lateral deflection during forward (trace) and reverse (retrace) scans and then calculating the difference. The lateral retrace deflection is opposite in sign to the lateral trace deflection. Therefore, only the deflection due to friction is retained after subtraction. The frictional data thus obtained is plotted as a function of the position of the tip on the surface in order to construct the frictional force image.

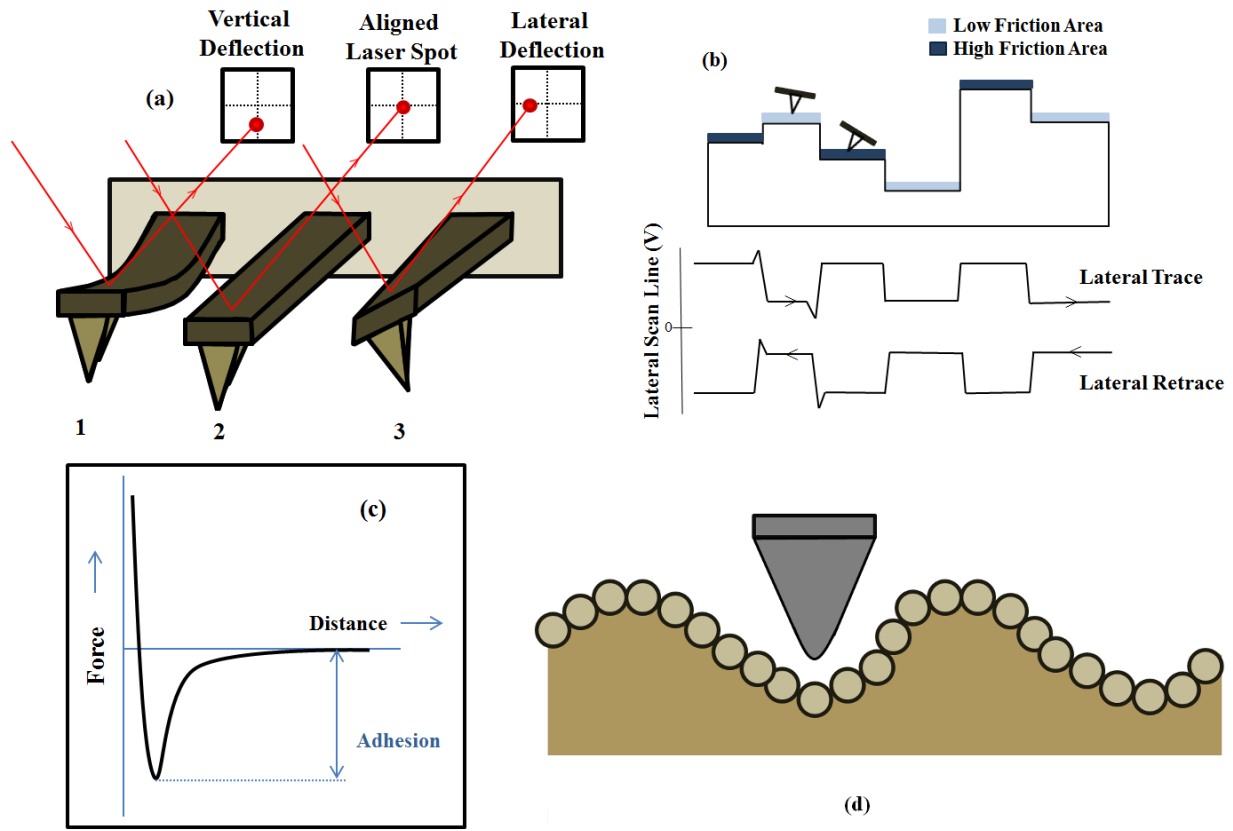

**Figure 1:** Schematic description of (a) The vertical and lateral cantilever deflection. (b) Frictional force measurement protocol (c) A typical force curve obtained for adhesion force measurements using an AFM cantilever. (d) A surface with low local asperity and high hills and valleys. The AFM tip interacts differently to these during friction and adhesion measurements.

In Figure 2(a) we demonstrate the frictional force distribution in an area of 50 μm$^2$ on the surface of the wing. In the bright regions the tip encounters higher friction as compared to the dark regions. From the visual inspection it is clear that the relative friction is higher in some elongated islands of length ~10 μm and width ~2 μm. The long axes of all such islands in the scanned area are oriented in the same direction. The separation between two nearest domains is ~ 10 μm. In Figure 2(b) we present the topographic image of the same area as in Figure 2(a). It is clear that the islands having higher friction also have higher topography. Therefore, the frictional force distribution on the wing surface seems to be strongly related with the topography. When we zoom into an area of 5um in the background, we observe ~1um wide grain-like structures (Figure 2(c)). These are not present on the high islands. On the other hand, on the high islands there are parallel nanowire-like structures of diameter ~ 5nm (Figure 2(d)). These wire-like structures are oriented perpendicular to the long axes of the islands. At certain locations thicker (diameter ~ 10 nm) ridge-structures run perpendicular to the 5nm diameter wires.

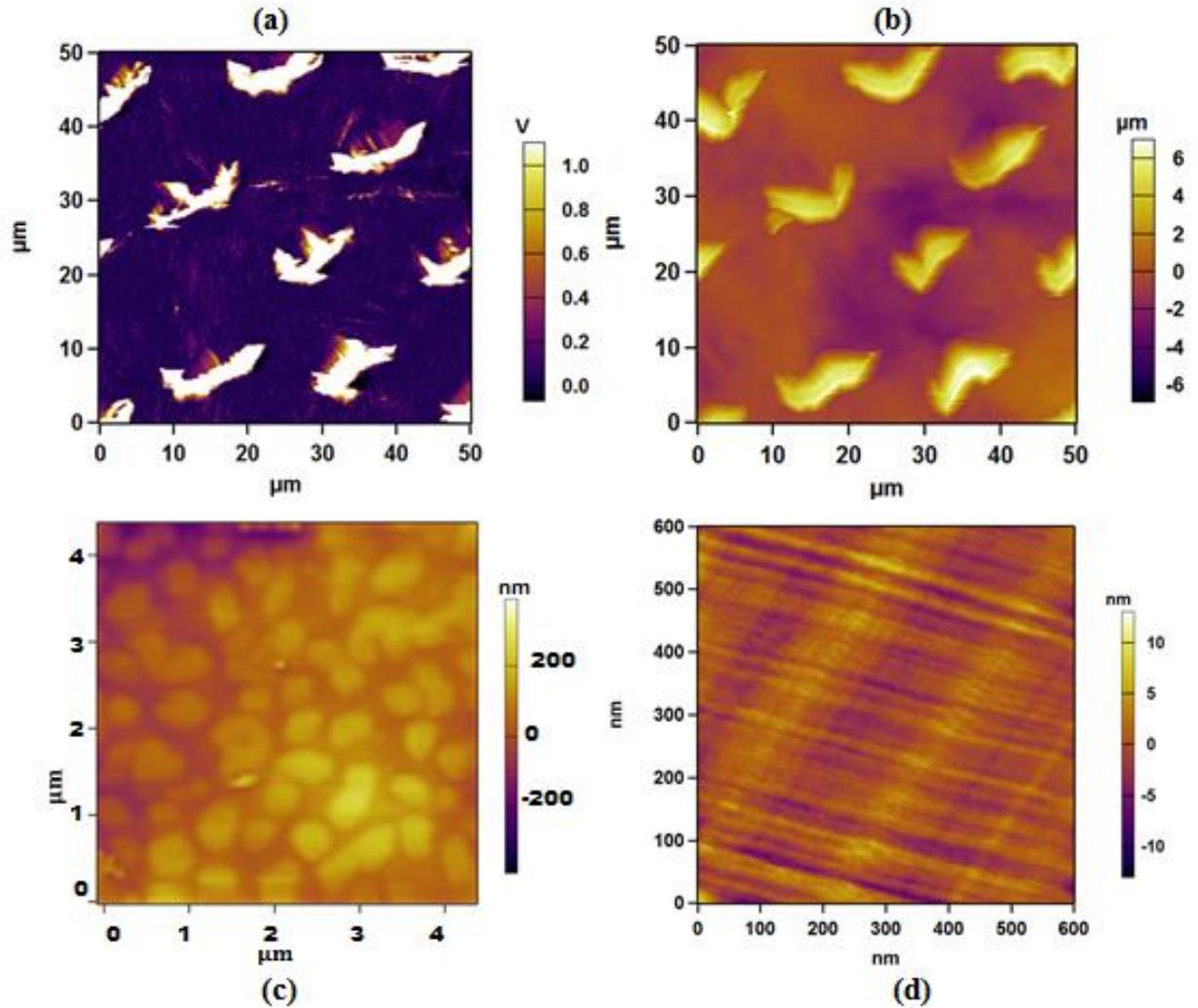

**Figure 2: (a) Distribution of relative frictional forces on a 50μm² area on the wing surface. (b) Topographic image of the same area as in (a) topographic height distribution. (c)Higher resolution topographic image of the back ground showing granular topography. (d) Nanostructures on top of one of the islands.**

One of the side-walls of each island is slanted and is decorated with the above-mentioned nanowire-like structures. Such nano-structuring might or might not be present in the opposite wall-- it is not clearly resolved due to the presence of a concave curvature in that wall and therefore the tip cannot probe the wall surface.

In order to investigate the adhesion properties of the wings, we have measured the distribution of the adhesive forces on the wing. For this measurement the wing is loaded flat on a glass slide and the measurement is done using a silicon cantilever coated with titanium and iridium and a spring constant of 2N/m. First the AFM tip is allowed to approach the surface with the feedback loop on. The approach is stopped as soon as the deflection of the cantilever reaches a pre-set value (the set point). From this point the tip is allowed to retract slowly and the force required to pull the tip out of the attractive regime of the force curve is measured which is a measure of the adhesion force. This could be approximated to be equal to the depth of the attractive region of the force vs. distance curve as shown in Figure 1(c). The force curve is recorded for a large number of points on the surface and the adhesion force plotted as function of the position of the AFM tip gives the adhesion image. It is observed that the adhesion force has a lower value on the island compared to the background, for all the islands that we measured. Figure 3(a) depicts the adhesion force map on a 16 µm² area including one island of higher friction. The force curve was recorded at 70×70 points on the given area for obtaining this plot. Figure 3(b) shows the topographic image of the single island on which the measurement for Figure 3(a) was done.

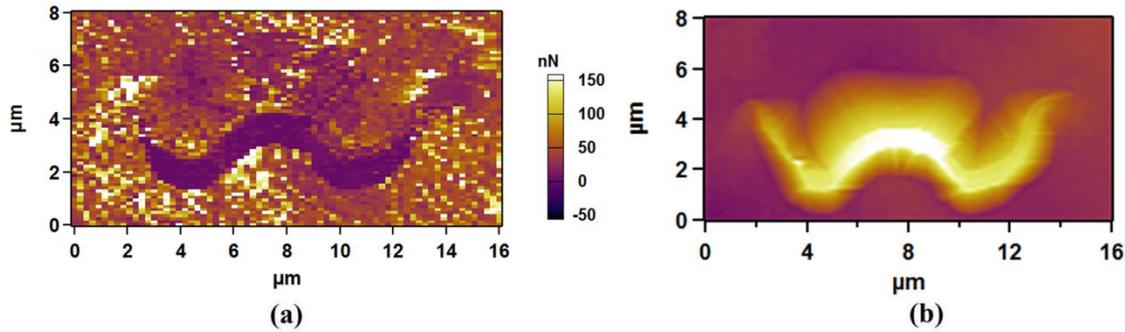

Figure 2: (a) Modulation of the adhesive forces in a 16µm² area containing one high island. (b) Non-contact AFM topographic image of the single island as in (a).

It should be noted that in the past adhesion force mapping was done on wings of drosophila melanogaster and it was found that the wings had nano-scale bumps on which the adhesion was not significantly different from the surrounding membrane [16]. The overall adhesion values on the wings previously found was fairly low [2,17-19].

From our measurements it is clear that the friction is higher but adhesion is lower on the islands compared to the background. It is common wisdom that friction is related to the roughness of the surface. From past theoretical studies it is also known that surface roughness may also influence adhesion properties. Within the theoretical model developed by Rumpf [20], the adhesion force between a particle and a surface is given by $F_{adh} = \frac{A_H R}{6 D_0} \left[ \frac{r}{r+R} + \frac{1}{(1+r/D_0)^2} \right]$, where R is particle radius, r is asperity radius on the surface, $A_H$ is Hamekar constant and $D_0$ is the minimum distance between sphere and asperity (interatomic spacing). If the asperity radius on the surface is small, the AFM tip will experience a lower adhesion. Comparing the structural details of Figure 2(c) and Figure 2(d) it is clear that the effective value of *r* is higher on the background than on the high islands: The mean asperity radius on the islands is around 5 nm while the bigger granular structures in the background may lead to a higher value of *r*. This explains the lower adhesion on the high islands compared to the background. However, friction between the AFM tip and the surface is also affected by the larger hills and valleys existing on a surface. As shown in Figure 1(d), the small circular structures on the sample around the tip will contribute to adhesion but the larger hill-like structures will not contribute for a given point on the surface where the adhesion is measured. However, the larger hill-like structures should contribute strongly to friction as for friction measurement the tip rubs laterally against the surface. The topography on the islands has several hills and valleys as seen in the image of a single island in Figure 3(b) leading to a higher friction on the valleys. Such modulation of topography does not exist in the background.

From the above measurements it is clear that the surface of the wings has well defined islands that are higher in height than the background having higher friction and lower adhesion than the background. It is known that lower adhesion of the wings give rise to their self-cleaning properties like in lotus leaves [3,17,20-28]. The specific architecture and enhanced friction might possibly help the insects get aerodynamic advantage during strong winds [29-30]. However, it is remarkable that such properties are localized at certain specific locations. It appears that additional volume is needed for required nanostructuring in order to felicitate the desired mechanical properties. However, additional volume leads to increase in mass of the wing which is a disadvantage for flying. Hence, we may surmise that the advantageous mechanical properties have been distributed over the wing such that the total mass is minimized without compromising on the advantages.

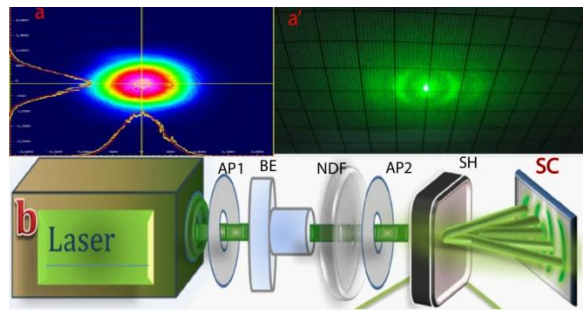

Figure 3: (a) Incident beam profiles for a green solid-state laser (λ = 523nm). (a') digital photograph of the diffraction pattern on the screen (b) schematic of the set-up with various components SH:wing sample hold ; AP1,AP2: iris; C: beam expander; SC: screen and NDF: ND filter wheel.

AFM meausrements can probe the distribution on a small area (max. 90umx90um in our case). In order to probe the distribution of the islands over the wing, we have performed optical diffraction experiments using a green solid-state laser of wavelength 523nm [31]. The experimental set up is shown in Figure 4. The laser spot (size = 2mm) was irradiated on the wing mounted on the path of the laser and the diffraction pattern was captured on a screen that was mounted 20.5cm away from the wing. The central bright spot and the higher order maxima were clearly visible in the recorded pattern as shown in Figure 4(a). Since the length-scale of the islands are comparable to the wavelength of the laser beam, the islands, if they are spatially correlated over a large area, would ideally behave like a grating generating the diffraction pattern. The pattern was stable and reproducible indicating the presence of a long-range spatial correlation of the islands in the wing.

To conclude, we have shown that the transparent insect wings of rain flies have microscopic islands on which the AFM tip experiences higher friction and lower adhesion. Optical diffraction experiments demonstrate that the islands are distributed over a long range in a spatially correlated manner. Therefore, the nano-scale mechanical properties of the wings are correlated with the topographic features over the entire wing. The unique arrangement and distribution of mechanical forces correlated with the topographic modulations might contribute to the aerodynamic advantage that the insects need during strong winds. Outstanding self-cleaning and anti-wetting properties of the wing might also originate from the superior nanostructuring on the wing-surface. The island formation might be motivated by the requirement of lower mass of the wing while retaining the necessary nano-mechanical advantages.

[*] **Corresponding author:** goutam@iisermohali.ac.in